\documentclass[aps,pra,twocolumn,showpacs,groupedaddress]{revtex4}
\usepackage{mathrsfs}
\usepackage{amsfonts}
\usepackage{amssymb}
\usepackage{graphicx}

\newcommand{\sn}{\text{sn}}
\newcommand{\cn}{\text{cn}}

\begin{document}
\title{Dynamics of quantum-classical hybrid system:
effect of matter-wave pressure}

\author{J. Shen$^1$, X. L. Huang$^1$, X. X. Yi$^{1,2}$\footnote{yixx@dlut.edu.cn},
Chunfeng  Wu$^2$, and C. H. Oh$^2$}
\affiliation{$^1$School of Physics and Optoelectronic Technology\\
Dalian University of Technology, Dalian 116024 China\\
$^2$Centre for Quantum Technologies and Department of Physics,
National University of Singapore, 117543, Singapore }

\date{\today}

\begin{abstract}
Radiation pressure  affects the kinetics of a system exposed to the
radiation and it constitutes  the basis of laser cooling. In this
paper, we study {\it matter-wave pressure} through examining the
dynamics of a quantum-classical hybrid system. The quantum and
classical subsystem have no explicit coupling to each other, but
affect mutually via a changing boundary condition. Two systems,
i.e., an atom and a Bose-Einstein condensate(BEC), are considered as
the quantum subsystems, while an oscillating wall is taken as the
classical subsystem. We show that the classical subsystem would
experience a force proportional to $Q^{-3}$ from the quantum atom,
whereas  it acquires  an additional force proportional to $Q^{-2}$
from the BEC due to the atom-atom interaction in the BEC.  These
forces can be understood  as the {\it matter-wave pressure}.
\end{abstract}

 \maketitle

\section{introduction}
It is well known  that electromagnetic radiation exerts a pressure
upon any surface exposed to it, the pressure  was deduced
theoretically by James Clerk Maxwell  and Adolfo Bartoli more than a
century ago, and proven experimentally by Lebedev, Ernest Fox
Nichols and Gordon Ferrie Hullin in the last century. Recently, this
research field becomes active\cite{A. Dorsel83, P. F. Cohadon99, J.
Kippenberg05, A. Schliesser06, Florian Marquardt06, M.
Bhattacharya07, D. Meiser06, D. Meiser062, M. Bhattacharya08} again
due to the breakthrough in nanofabrication and in ultracold science,
this together with the coupling of coherent optical system to
micromechanical devices, has opened up the exciting new field of
research, cavity optomechanics\cite{D. Kleckner06, O. Arcizet06,
T.Corbitt07, K. Hammerer, Florian Marquardt}.

Quantum mechanics tells us that  all matter exhibits both wave-like
and particle-like properties, called wave-particle duality, the
wavelength is inversely proportional to the momentum of a particle
and  the frequency is directly proportional to the particle energy.
These facts give rise to a question: Can the matter wave exert a
pressure ({\it matter-wave pressure}) upon a surface exposed to it,
like the electromagnetic radiation does? To answer this question, it
is good to study a coupled quantum-classical hybrid system, and
examine how the quantum subsystem affects the kinetics of the
classical subsystem.

On the other hand, classical-quantum hybrid system on its own  is an
interesting topic to study\cite{Biao Wu06,yi07}. By
classical-quantum hybrid system we mean a composite system
consisting of a quantum and a classical subsystem. For a closed
hybrid system, the quantum subsystem may be treated classically for
specific issues addressed \cite{C. P. Slichter90, G. P. Berman02}.
However, this approach  in general is inadequate  owing to the loss
of quantum features. In Ref.\cite{Biao Wu06}, the authors present a
general  framework for  exact treatment of such a hybrid system.
When the quantum subsystem is dynamically fast and the classical
subsystem is slow, a vector potential arises.  This vector
potential, on one hand, gives rise to the familiar Berry phase in
the fast quantum dynamics, on the other hand, it yields a
Lorentz-like force in the slow classical dynamics. In the formalism,
the Hamiltonian of the fast quantum subsystem depends explicitly on
the freedom  of the classical subsystem. In contrast, here we  study
the dynamics   of quantum-classical hybrid system
 without any explicit  interaction, instead an
boundary condition will be considered that connects the quantum and
the classical subsystem. An atom and a BEC are taken  as the quantum
subsystem, both of them move in an one-dimensional square well with
an oscillating wall as their boundary.   The difference between
these two quantum subsystems is that the BEC has atom-atom
interaction, whereas the atom would not have. This setting is
exactly a model we need to discuss the problem of {\it matter-wave
pressure}.

The paper is organized as follows. After a brief introduction to the
formalism for classical-quantum hybrid system in Sec.{\rm II}, we
study the kinetics  of the classical subsystem (i.e., the
oscillating wall) under the effect of a quantum atom in Sec.{\rm
III}. Treating the atom as a classical particle, the kinetics of the
wall is also examined in this section.  In Sec.{\rm IV}, we
investigate  the kinetics of the wall exposed  to a Bose-Einstein
condensate. The atom-atom coupling in the BEC will make the force
acting on the wall different from the case in Sec. {\rm III}.
Finally we conclude our results in Sec.{\rm V}.

\section{general formalism}
In this section, we present a brief introduction to the formalism
for quantum-classical hybrid system. For more detail, we refer the
reader to Ref.\cite{Biao Wu06}.  Consider a quantum-classical hybrid
system described by
\begin{eqnarray}
H=\langle\mathbf{\Psi}|\hat{H}_{1}(\mathbf{Q})|\mathbf{\Psi}\rangle
+H_{2}(\mathbf{P},\mathbf{Q}),
\end{eqnarray}
where $\hat{H}_{1}$ is the Hamiltonian for the quantum subsystem and
$|\mathbf{\Psi}\rangle=(\psi_{1},\psi_{2},...,\psi_{N})^T$ denotes
its quantum state.  $H_{2}(\mathbf{P},\mathbf{Q})$ represents  the
classical subsystem and $\mathbf{P}$, $\mathbf{Q}$ stand for its
coordinates and momenta, respectively. It has been  shown\cite{A.
Heslot85, S. weinberg89} that a quantum system possesses
mathematically a canonical classical Hamiltonian structure. This can
be understood by defining $x_{i}=\sqrt{i\hbar}\Psi_{i}$,
$y_{i}=\sqrt{i\hbar}\Psi^{*}_{i}$. With this definition, one can
rewrite the dynamics for the quantum system as
\begin{eqnarray}
&&\frac{dx_{i}}{dt}=\frac{\partial H^{'}_{1}}{\partial y_{i}},\nonumber\\
&&\frac{dy_{i}}{dt}=-\frac{\partial H^{'}_{1}}{\partial x_{i}},
\label{totalequation}
\end{eqnarray}
where $H^{'}_{1}=H^{'}_{1}(y_{i}, x_{i}, \mathbf{Q})=
H_{1}(\mathbf{\Psi},\mathbf{\Psi^{*}},\mathbf
{Q})=\langle\mathbf{\Psi}|\hat{H}_{1}(\mathbf{Q})|\mathbf{\Psi}\rangle.$
Thus the quantum system can be reformulated in the language of
classical theory Eq.(\ref{totalequation}).

The quantum state $|\mathbf{\Psi}\rangle$ can be expanded in terms
of instantaneous eigenstates of the Hamiltonian $\hat{H}_1$,
\begin{equation}
|\mathbf{\Psi}\rangle=\sum_{n=1}^{N} c_{n}|\phi_{n}(\mathbf
Q)\rangle,
\end{equation}
where $\hat{H}_{1}(\mathbf Q)|\phi_{n}(\mathbf
Q)\rangle=E_{n}(\mathbf Q)|\phi_{n}(\mathbf Q)\rangle$. By the
formalism in Ref.\cite{Biao Wu06}, we can obtain a vector potential
that the classical subsystem feels,
\begin{equation}
\mathbf{A}=\sum_{n=1}^{N} I_{1n}\mathbf{A}_{n},\qquad
\mathbf{A}_{n}=i\langle\phi_{n}|\frac{\partial}{\partial
\mathbf{Q}}|\phi_{n}\rangle,
\end{equation}
where $I_{1n}=\hbar|c_{n}|^{2}$. This leads to a dynamical equation
for the classical subsystem,
\begin{equation}
M\bar{\ddot{\mathbf{Q}}}=-\frac{\partial\mathscr{H}_{1}}{\partial
\mathbf{Q}}-\frac{\partial V_{2}}{\partial
\mathbf{Q}}+\bar{\dot{\mathbf{Q}}}\times\mathbf{\mathscr{B}},\label{classicalequation}
\end{equation}
with $\mathscr{H}_{1}=\mathscr{H}_{1}(\mathbf{I}_{1}, \mathbf
{Q})=\sum_{n}E_{n}(\mathbf{Q})I_{1n}/\hbar,$ $V_{2}$ is a potential
and
$\mathbf{\mathscr{B}}=\nabla\times\mathbf{A}=\sum_{n}I_{1n}\nabla\times\mathbf{A}_{n}$
is a gauge field  like  magnetic field.

\section{atom as the quantum subsystem}
\begin{figure}
\includegraphics*[width=0.6\columnwidth,
height=0.5\columnwidth]{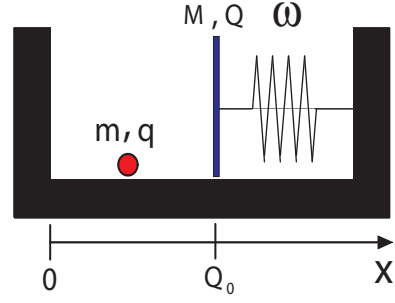} \caption{Schematic diagram
of the quantum-classical hybrid system.} \label{model}
\end{figure}

Consider a bipartite hybrid system which consists  of a single atom
in an infinitely  deep well  and a moving wall, the moving wall acts
as a boundary for the atom (see Fig.\ref{model}), and the whole
system is restricted to move in one-dimension.
The Hamiltonian of such a system can be described by
\begin{eqnarray}
\hat{H}=\hat{H}_{1}+H_{2},\label{totalHamiltonian}
\end{eqnarray}
where $\hat{H}_{1}$ denotes the Hamiltonian of the atom trapped in
the infinitely deep well, $H_{2}$ stands for the Hamiltonian of the
moving wall which is considered as a classical harmonic oscillator.
Here $\hat{H}_{1}=p^2/2m$, and
$H_{2}=\frac{P^2}{2M}+\frac{1}{2}M\omega^{2}(Q-Q_{0})^{2}$. $p$
 and $P$ denote the momentum along the $x$ axis for the quantum atom
and classical wall respectively; Note $p$ is an operator while $P$
is a c-number  here; $m$ and $M$ stand for the mass of the atom and
the oscillator respectively;
$V_2=\frac{1}{2}M\omega^{2}(Q-Q_{0})^{2}$ is the potential energy of
the harmonic oscillator; $Q$ is the coordinate of the moving wall
and $Q_{0}$ is its  equilibrium position; $\omega$ denotes the
vibration frequency of the classical subsystem. In this hybrid
system, we consider a situation that there is no explicit
interaction between the quantum atom and the moving wall. The moving
wall affects the dynamics of the quantum atom by changing its
boundary condition. Interesting features arise in this case as we
will show below.

In our model the classical subsystem only changes the boundary
condition of the quantum subsystem, which would reflect in  the
instantaneous eigenstate and the corresponding  eigenvalue of the
quantum subsystem given below
\begin{eqnarray}
&&|\phi_{n}(Q)\rangle=\int_{0}^{Q}\phi_{n}(q, Q)|q\rangle dq, \nonumber\\
&&\phi_{n}(q, Q)=\sqrt{\frac{2}{Q}}\sin\frac{n\pi q}{Q}, \nonumber\\
&&E_{n}(Q)=\frac{\hbar^{2}\pi^{2}n^{2}}{2mQ^{2}},
\end{eqnarray}
where $|q\rangle$ denotes the eigenstate of coordinate $q$. Then the
Hamiltonian of the quantum subsystem can be rewritten  as
\begin{equation}
\mathscr{H}_{1}(\mathbf{I_{1}},Q)=\sum_{n}E_{n}(
Q)I_{1n}/\hbar=\sum_{n}\frac{n^{2}\pi^{2}\hbar^{2}}{2mQ^{2}}|c_{n}|^{2}.
\end{equation}
With these results, the vector potential $\mathbf{A}$ is then
\begin{eqnarray}
\mathbf{A}&=&\sum_{n=1}^{N} I_{1n}\mathbf{A}_{n}\nonumber\\
&=&\sum_{n=1}^{\infty}i\hbar|c_{n}|^{2}\int_{0}^{Q}\phi_{n}^{*}(q, Q)
\frac{\partial}{\partial Q}\phi_{n}(q, Q) dq\nonumber\\
&=&0,
\end{eqnarray}
 leading to an observation that only scalar potential (from the
 quantum subsystem) affects the dynamics of
the classical subsystem. It is worth noting  that this conclusion
depends on the model, i.e., the magnetic-like gauge fields are not
zero in general.

Substituting these equations together with the potential energy
$V_{2}$ into  Eq.(\ref{classicalequation}), we  get the dynamical
equation for the classical subsystem of the linear system
(one-dimensional)
\begin{equation}
{\ddot{Q}}=\frac{B}{Q^{3}}-\omega^{2}(Q-Q_{0})\label{classicalequation2},
\end{equation}
where $B=\sum_n\frac{n^2\pi^{2}\hbar^{2}}{mM}$. This equation
includes the effect of the quantum subsystem and describes the
kinetics  of the classical subsystem. The first term in
Eq.(\ref{classicalequation2}) represents a force acting on the wall
from the atom. We explain this force as a consequence of matter-wave
pressure for the following reasons. (1) Suppose that the population
of the atom in level $n$ remains unchanged, the total energy of the
atom is
$\bar{E}_1=\sum_{n}\frac{n^{2}\pi^{2}\hbar^{2}}{2mQ^{2}}|c_{n}|^{2},$
where $|c_n|^2$ is the probability of the atom in level $n$.
Consider a small vibration $\Delta Q$ of the wall, the work done by
the atom is $\bar{F}\cdot \Delta Q$, and we have
$\bar{F}=-\frac{\partial \bar{E}_1}{\partial Q}$, this exactly leads
to the first term in Eq.(\ref{classicalequation2}). (2) The
wave-function of the atom is $|\Phi\rangle=\sum_n
c_n|\phi_n(Q)\rangle,$ with the energy $\bar{E}_1$ and momentum
$\bar{p}=\sum_n|c_n|^2 p_n=\sum_n|c_n|^2 \frac{n\pi \hbar}{Q},$ the
probability current is $J=\sum_n J_n=\sum_n |c_n|^2\frac
{p_n}{2mQ}.$ This means that the atom exerts a pressure force given
by $\bar{F}=\sum_n 2p_n
J_n=\sum_n|c_n|^2\frac{n^2\pi^2\hbar^2}{mQ^3}$ on the wall.
\begin{figure}
\includegraphics*[width=0.8\columnwidth,height=0.5\columnwidth]{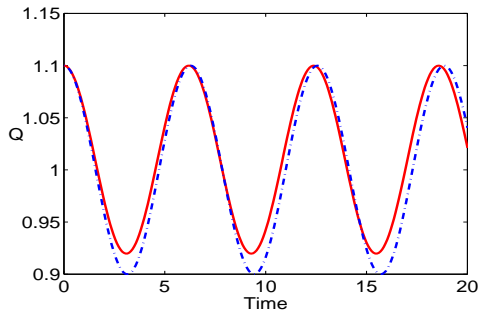}
\caption{The coordinate of the classical subsystem $Q$ (in units of
$nm$) as a function of time (in units of $10^{-7}$s) with $B=0.01$,
$\omega=1$ (in units of $2\pi\times 10^7$ Hz), $Q_{0}=1$, the
initial condition is $Q(0)=1.1$ and $\dot{Q}(0)=0$. In the red solid
curve, we consider the effect of the quantum subsystem, while in the
blue dash curve we plot the free evolution of the oscillator.}
\label{Fig1}
\end{figure}
\begin{figure}
\includegraphics*[width=0.8\columnwidth,height=0.5\columnwidth]{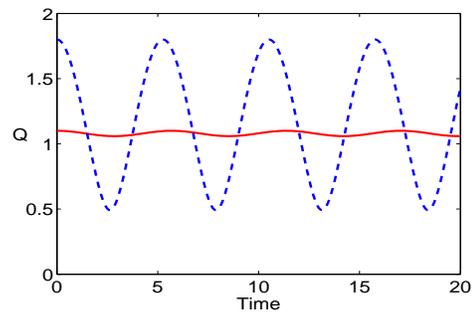}
\caption{(Color online) The time evolution for the coordinate of the
classical subsystem $Q$ under  different initial amplitudes
($Q(0)-Q_{0}$) for the parameters of $Q(0)=1.1$ (red solid curve)
and $Q(0)=1.8$ (blue dash curve). The other parameters are chosen as
$B=0.1$, $\omega=1$, $\dot{Q}(0)=0$ and $Q_{0}=1$. The frequency,
time and length were rescaled as the same as  Fig.\ref{Fig1}.}
\label{Fig2}
\end{figure}

\begin{figure}
\includegraphics*[width=0.8\columnwidth,height=0.5\columnwidth]{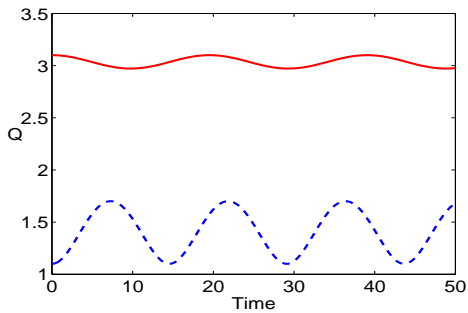}
\caption{(Color online) The time evolution of  the coordinate  $Q$
for  different initial equilibrium positions  $Q_{0}$ but the same
initial amplitude ($Q(0)-Q_{0}=0.1$). Here the parameters are chosen
as $B=\omega=0.1$, $\dot{Q}(0)=0$.  The red solid curve is plotted
for $Q_{0}=3,$  while for the blue dash curve, $Q_{0}=1$.}
\label{Fig3}
\end{figure}

In Fig.\ref{Fig1}, we plot the coordinate of the classical subsystem
$Q$ as a function of time. It is clear that $Q$ is a oscillating
function of time. The blue-dash line denotes a free harmonic
oscillator and the red-solid line is for the case with the
 effect of the quantum subsystem. Here the parameters are chosen as $B=0.01$,
$\omega=1$, $Q_{0}=1$， $Q(0)=1.1$ and $\dot{Q}(0)=0$, where $Q(0)$
and $\dot{Q}(0)$ denote the initial maximal displacement from the
equilibrium  and the initial velocity of the moving wall,
respectively. Comparing with the free harmonic oscillator (the blue
dash line in Fig.\ref{Fig1}),  we find that when we consider the
effect of quantum subsystem the equilibrium point moves to the right
side. This is a consequence of the matter-wave pressure. In
Fig.\ref{Fig2}, we present the time evolution of the coordinate of
the classical subsystem $Q$ under  different initial amplitudes
($Q(0)-Q_{0}$). For the red solid curve, we choose $Q(0)=1.1,$ while
for the blue dash curve, $Q(0)=1.8$. The other parameters are
$B=0.1$, $\omega=1$, and $Q_{0}=1$. We find that the effect of the
quantum subsystem increases with  $Q(0)-Q_{0}$. The curve becomes
sharper when the coordinate moves toward the origin. This feature
becomes evident   for the case  of large initial amplitude. Next we
show the coordinate $Q$ as a function of time with different initial
equilibrium position  $Q_{0}$ in Fig.\ref{Fig3}. We find that for
the same initial amplitude, the larger initial equilibrium position
$Q_{0}$ is, the smaller the effect of quantum subsystem on the
classical subsystem.

These observations  can be understood  by examining  the dynamical
equation Eq.(\ref{classicalequation2}). In this equation
$M{\ddot{Q}}$ denotes the  resultant force of the moving wall and
$M\omega^{2}(Q-Q_{0})$ is the elastic spring force which keeps the
moving wall in harmonic oscillating. Quantum subsystem brings in a
rightward  force ($\frac{B_0}{Q^{3}}$, with $B_0=BM$) which is
inversely proportional to the cube of the coordinate $Q$ of the
moving wall. This is the reason why the equilibrium position moves
more to the right side than the free harmonic oscillator. The
quantum subsystem induced force results from the matter-wave
pressure, hence by examining the kinetics of the classical
subsystem, we may recognize  some features of the matter-wave
pressure.   It is easy to see that this force decreases  rapidly
with the increasing of $Q$. This is the reason why the effect of the
quantum subsystem become smaller when the width of the well gets
larger. In other words, this force changes in a wide range when the
amplitude of the oscillation (of the wall) is large.
\begin{figure}
\includegraphics*[width=0.8\columnwidth,height=0.5\columnwidth]{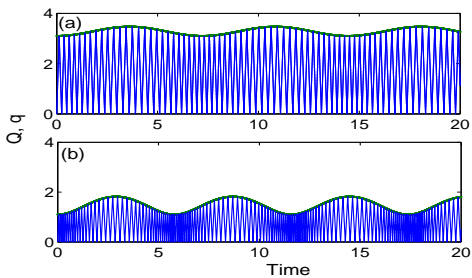}
\caption{(Color online) In contrast to Fig.\ref{Fig3}, in which that
atom was treated as a quantum system, in this figure we show the
coordinate $Q$ (thick green) and $q$ (thin
 blue) as a function of time. Both the atom and the wall are
considered as classical objects. This plot is for different initial
equilibrium positions $Q_{0}$, but the same initial amplitude
($Q(0)-Q_{0}=0.1$), (a) $Q_0=3,$ (b) $Q_0=1$. The other parameters
chosen are $M\omega^2=60,$ $M/m=1000$, the initial velocity of the
atom and the wall are  $25$ and $0$, respectively. The atom was
assumed initially at the origin. In this plot, time is in units of
$10^{-7}$s, length in units of nm. } \label{Fig4}
\end{figure}

It is interesting to compare these observations with the results
given by classical mechanics. To this end, we simulate numerically
the kinetics of the hybrid system, treating both the atom and the
wall classically. The Hamiltonian that governs the kinetics of such
a classical system is the same as Eq.(\ref{totalHamiltonian}), but
$p$ in $\hat{H}_1$ is not an operator now. The atom is bouncing back
and forth and colliding with the  wall. We assume that the collision
is elastic such that both energy and momentum are conserved. The
numerical results are presented in Fig.\ref{Fig4}, where we plot the
coordinates of the atom ($q$) and wall ($Q$) as a function of time.
Similar to the quantum case, the smaller the well is (or,  the
smaller the $Q_0$), the lager the effect of the atom on the wall.
The difference is that a quantum atom exert a force proportional to
$1/Q^3$ on the wall, while a classical atom provides a force on
average  proportional to $1/Q$ according to the law of conservation
of momentum. As a consequence, the kinetics of the wall behaves
differently for $Q\rightarrow 0.$

\section{Bose-Einstein condensate as the quantum subsystem}

In this section we will take a BEC as the quantum subsystem to study
the kinetics of the classical subsystem. The difference between the
atom and the BEC that  will manifest in our study is the atom-atom
interaction  in the BEC. We will show in the following that this
atom-atom interaction results in  an additional force proportional
to $Q^{-2}$ to the classical subsystem. For simplicity, we consider
a BEC in one-dimensional square well with an moving wall as its
boundary. The setting is the same as in Fig.\ref{model}, but the
atom is replaced with a BEC. The Hamiltonian which describes such a
system can be written as,
\begin{eqnarray}
H=\int{\psi}^{*}(x)\hat{H}_{1}(x,Q)\psi(x)\text{d}^{3}x +H_{2}(P,Q),
\end{eqnarray}
where  $\hat{H}_{1}$ denotes the Hamiltonian for the BEC, and
$\psi(x)$ is the wave function of the BEC. $H_{2}(P, Q)$ is the
Hamiltonian of the classical moving wall and $P$, $Q$ are its
coordinate and momentum, respectively. The effective Hamiltonian of
a BEC in a potential $V(x)$ takes,
\begin{eqnarray}
\hat{H}_{1}=-\frac{\hbar^{2}}{2m}\frac{\partial^{2}}{\partial
x^{2}}+V(x)+g|\psi(x,t)|^{2}, \label{GPequation}
\end{eqnarray}
where the potential $V(x)$ in our model is,
\begin{eqnarray}
V(x)=\left\{\begin{array}{cc}
       0 & 0\leq x \leq Q \\
       +\infty &  x<0 ~\text{or} ~x>Q
     \end{array}\right. .
\end{eqnarray}
Here $x$ stands for  the coordinate of the BEC and the atom-atom
coupling constant in BEC is denoted by $g$. The stationary GP
equation  can be written as
\begin{eqnarray}
\left(-\frac{\hbar^{2}}{2m}\frac{\partial^2}{\partial x^2}
+V(x)+g|\psi(x,t)|^{2}\right)\psi(x,t)=u\psi(x,t).\nonumber\\
\label{GPequation2}
\end{eqnarray}
Here $u$ denotes the chemical potential. In terms of Jacobi elliptic
functions, the eigenfunction for repulsive interaction $g>0$ and
attractive interaction  $g<0$  can be written as\cite{W. D. Li06, L.
D. Carr00},
\begin{eqnarray}
&&\varphi_{+,j}(x)=b_{+, j}\text{sn}(a_{+,j}\cdot x
+\delta_{+, j}, k_{+, j}), \quad \text{for} ~g>0, \nonumber\\
&&\varphi_{-,j}(x)=b_{-, j}\text{cn}(a_{-, j}\cdot x
+\delta_{-, j}, k_{-, j}), \quad \text{for} ~g<0,\nonumber\\
\label{GPsolution1}
\end{eqnarray}
with
\begin{eqnarray}
k_{\pm, j}=\frac{b^{2}_{\pm,j}}{2a^{2}_{\pm, j}}|g|,
\end{eqnarray}
where $\sn$, $\cn$ are the Jacobi elliptic functions and $k_{\pm,j}\
\ (j=1,2,3,...)$ are the modular number of the Jacobi elliptic
function, $\delta_{\pm,j}\ \ (j=1,2,3,...)$ are  constants which
will be given below. We use $+$ and $-$ to denote the case of $g >
0$ and $g < 0$ respectively, and $j$ labels the eigenfunctions.

Taking  the boundary conditions
$\varphi_{\pm,j}(0)=\varphi_{\pm,j}(Q)=0$ and the normalization
condition $\int_{0}^{Q}|\varphi_{\pm,j}(x)|^{2}dx=1$ into account,
we obtain,
\begin{eqnarray}
&&\delta_{+,j}=0,\nonumber\\
&&a_{+, j}=\frac{2jK(k_{+, j})}{Q},~j=1,2,3...\nonumber\\
&&b_{+, j}=\sqrt{\frac{k_{+, j}K(k_{+, j})}{[K(k_{+, j})-E(k_{+, j})]Q}},\nonumber\\
&&\delta_{-,j}=-K(k_{-,j}),\nonumber\\
&&a_{-,j}=\frac{2jK(k_{-,j})}{Q},~j=1,2,3...\nonumber\\
&&b_{-,j}=\sqrt{\frac{k_{-,j}K(k_{-,j})}{[E(k_{-,j})+(k_{-,j}-1)K(k_{-, j})]Q}},
\end{eqnarray}
where  $K(k_{\pm,j})$ and $E(k_{\pm,j})$ are the first and the
second elliptic integrals, respectively.

We now focus on  the kinetic equation of the classical subsystem. In
our model, it is easy to show that
$\mathbf{\mathscr{B}}=\nabla\times\mathbf{A}=0$, and the classical
Hamiltonian for the BEC takes
\begin{eqnarray}
\mathscr{H}_{1}
=&&\int_{0}^{Q}\psi^{*}(x,t)\large[-\frac{\hbar^{2}}{2m}\frac{\partial^{2}}{\partial
x^{2}}+V(x)\nonumber\\&&+\frac{1}{2}g\psi^{*}(x,t)\psi(x,t)\large]
\psi(x,t)dx\label{Hamiltionnonlinear}.
\end{eqnarray}
\begin{widetext}
Specifically, for $g>0$
\begin{eqnarray}
\mathscr{H}_{1, +, j}=&&\frac{\hbar^{2}}{2m}\frac{2jk_{+,
j}K^2(k_{+, j})} {Q^{2}[K(k_{+, j})-E(k_{+, j})]} \int^{2jK(k_{+,
j})}_{0}[(k_{+, j}+1)\sn^{2}z -2k_{+, j}\sn^4z
]dz\nonumber\\&&+\frac{gk_{+, j}^{2}K(k_{+, j})} {4j[K(k_{+,
j})-E(k_{+, j})]^{2}Q}\times\int^{2jK(k_{+, j})}_{0}\sn^{4}z dz,
\end{eqnarray}
and for $g<0$
\begin{eqnarray}
\mathscr{H}_{1,-, j}=&&\frac{\hbar^{2}}{2m}\frac{2jk_{-, j}
K^2(k_{-, j})}{Q^{2}[(k_{-, j}-1)K(k_{-, j})+E(k_{-, j})]}
\int^{(2j-1)K(k_{-, j})}_{-K(k_{-, j})}
[1-(2k_{-, j}+1)\sn^{2}y+2k_{-, j}\sn^{4}y]dy\nonumber\\
&&+\frac{gk_{-, j}^{2}K(k_{-, j})}{4j[(k_{-, j}-1)K(k_{-, j})
+E(k_{-, j})]^{2}Q} \times\int^{(2j-1)K(k_{-, j})}_{-K(k_{-,
j})}\cn^{4}y dy.~~~~~~~(g<0)
\end{eqnarray}
Here $z=\frac{2jK(k_{+, j})x}{Q}$ and $y=K(k_{-,
j})(\frac{2jx}{Q}-1)$. Substituting these equations together with
potential $V=\frac{1}{2}M\omega^{2}(Q-Q_{0})^{2}$ into
Eq.(\ref{classicalequation}), we obtain a kinetical  equation for
the classical subsystem. For repulsive interaction, i.e.,  $g>0$, it
is
\begin{eqnarray}
{\ddot{Q}}=\frac{C_{1}}{Q^{3}}+\frac{D_{1}}{Q^{2}}
-\omega^{2}(Q-Q_{0})\label{nonlineardynamics1},
\end{eqnarray}
with
$$C_{1}=\frac{2j\hbar^{2}k_{+, j}K(k_{+, j})^{2}} {Mm[K(k_{+,
j})-E(k_{+, j})]}\int^{2jK(k_{+, j})}_{0}[(k_{+, j}+1)\sn^{2}z
-2k_{+, j}\sn^4z ]dz>0,$$
 and
$$D_{1}=\frac{gk_{+, j}^{2}K(k_{+,
j})}{2jM[K(k_{+, j})-E(k_{+, j})]^{2}} \int^{2jK(k_{+,
j})}_{0}\sn^{4}z dz>0.$$
For $g<0,$
\begin{eqnarray}
{\ddot{Q}}=\frac{C_{2}}{Q^{3}}+\frac{D_{2}}{Q^{2}}
-\omega^{2}(Q-Q_{0})\label{nonlineardynamics2},
\end{eqnarray}
where
$$C_{2}=\frac{2j\hbar^{2}k_{-, j}K(k_{-, j})^{2}} {Mm[(k_{-,
j}-1)K(k_{-, j})+E(k_{-, j})]}\int^{(2j-1)K(k_{-, j})}_{-(k_{-, j})}
[1-(2k_{-, j}+1)\sn^{2}y+2k_{-, j}\sn^{4}y]dy>0,$$
 and
$$D_{2}=\frac{gk_{-, j}^{2}K(k_{-, j})}{2jM[(k_{-, j}-1)K(k_{-,
j})+E(k_{-, j})]^{2}} \int^{(2j-1)K(k_{-, j})}_{-K(k_{-,
j})}\cn^{4}y dy<0.$$
\end{widetext}
 It is easy to show that if there is no
interaction between the atoms  in BEC (namely, $g=0$), $C_i$ and
$D_i$ ($i=1,2$) reduce to
$C_{1}=C_{2}=\frac{j^{2}\pi^{2}\hbar^{2}}{Mm},$ and $
D_{1}=D_{2}=0$. This is exactly the result given in Sec.{\rm III},
where the atom was taken as the quantum subsystem.  Observing
Eqs.(\ref{nonlineardynamics1}) and (\ref{nonlineardynamics2}), we
find that the classical wall experiences  a force proportional to
$Q^{-3}$, which is the same as that we discussed in Sec.{\rm III}.
In addition to this force,   a force ($\sim Q^{-2}$)  inversely
proportional to the square of the coordinate of the classical
subsystem appears. This force is due to  the interaction between
atoms  in the BEC, which is different from the result we discussed
in Sec.{\rm III}.
\begin{figure}
\includegraphics*[width=0.8\columnwidth,height=0.5\columnwidth]{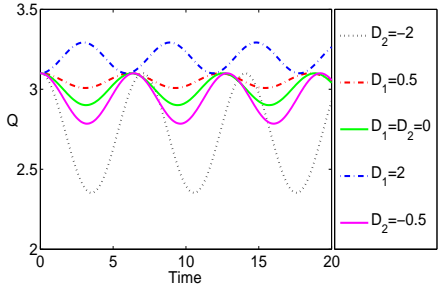}
\caption{(Color online) The time evolution of  the coordinate  $Q$
(in units of $nm$) for  different atom-atom interactions in the BEC.
The parameters and initial conditions are chosen as
$C_{1}=C_{2}=0.01$, $\omega=1$ (in units of $2\pi\times 10^7$ Hz),
the equilibrium position is $Q_{0}=3$, the initial position is
$Q(0)=3.1,$ and $\dot{Q}(0)=0$.} \label{Fig5}
\end{figure}

In Fig.\ref{Fig5} we plot the coordinate of the classical subsystem
$Q$ as a function of time for different $D_{1}$ and $D_{2}$. Here we
choose the parameters as $C_{1}=C_{2}=0.01$, $\omega=1$, $Q_{0}=3$,
$Q(0)=3.1$ and $\dot{Q}(0)=0$. In contrast, we plot the situation
without atom-atom interaction as the green solid line. From the
figure,  we find that the equilibrium point of the classical
subsystem moves right  for the case of $D_{1}=2$ and $D_{1}=0.5,$
whereas for the case of $D_{2}=-2$ and $D_{2}=-0.5,$ the equilibrium
 position  moves left.
 These   can be understood  by analyzing  the dynamical
 equation Eqs.(\ref{nonlineardynamics1}) and (\ref{nonlineardynamics2}).
 In Eq.(\ref{nonlineardynamics1}) $M{\ddot{Q}}$ is the  resultant
 force of the moving wall and
 $M\omega^{2}(Q-Q_{0})$ denotes the elastic spring force which keeps the
 moving wall in harmonic oscillation. Quantum subsystem brings in a
 rightward  force ($\frac{C_{1}}{Q^{3}}$)  similar to the case
 in Sec.{\rm III}.  Another term that is
 inversely proportional to the square of the coordinate($\frac{D_{1}}{Q^{2}}$)
comes from  the repulsive interaction between  atoms in the BEC. It
 results in  a repulsive force for  the moving wall, too. There
together can explain why the equilibrium position of the wall moves
right with respect to  the case without atom-atom coupling. For
similar  reasons the equilibrium position of the wall  moves  left
for attractive atom-atom interaction.

\section{Conclusion and discussions}
The dynamics of quantum-classical hybrid system has been studied in
this paper. Two quantum subsystems are taken to discuss the kinetics
of the classical subsystem.   When the quantum subsystem is an atom,
the classical subsystem experiences a  force ($\frac{B_0}{Q^{3}}$)
proportional to its distance $Q$  to the fixed wall, meanwhile the
energy of the atom has been changed  because the classical subsystem
provides a moving boundary, even if the atom remains in the same
level in  time evolution. When the quantum subsystem is a BEC, the
BEC would exert an additional force proportional to $Q^{-2}$ to the
moving wall. This force comes from the atom-atom interaction in the
BEC, hence it can act as  a witness of the nonlinearity in BEC. With
current technology, a SiN membrane of effective mass $M=4 \times
10^{-13}$Kg is possible in laboratory, a vibration frequency
$2\pi\times 1.3$ MHz sets the time scale of the dynamics to $\sim
10^{-7}$s. $m\sim 10^{-27}$Kg, $M\sim 10^{-13}$Kg, and the
eigenstate index $n\sim 500$ may lead to $B\sim 0.1$(in units of
$(Hz)^2m^4$) in the first example, with which the kinetics of the
classical system has been changed sharply. This estimation is
conservative. In fact, the frequency of the membrane can be $\sim$
GHz, leading to a time scale $\sim 10^{-9}$s. To our best knowledge,
this is the first time to show theoretically the effect of
matter-wave pressure, though the experimental observation is a
challenge task.
\ \ \\

We thank Biao Wu for  suggestion and comments.  This work is
supported by NSF of China under grant Nos 10775023 and 10935010, as
well as the National Research Foundation and Ministry of Education,
Singapore under academic research grant No. WBS:
R-710-000-008-271.\\

\end{document}